%% file: main.tex
\documentclass[conference]{IEEEtran}

\IEEEoverridecommandlockouts
\usepackage{cite}
\usepackage{amsmath,amssymb,amsfonts}
\usepackage{graphicx}
\usepackage{textcomp}
\usepackage{xcolor}
\def\BibTeX{{\rm B\kern-.05em{\sc i\kern-.025em b}\kern-.08em
    T\kern-.1667em\lower.7ex\hbox{E}\kern-.125emX}}

\usepackage{booktabs}
\usepackage{tabularx}
\usepackage[font=footnotesize]{caption}
\usepackage[font=footnotesize]{subcaption}

\usepackage{algorithm}
\usepackage{algpseudocode}
\usepackage{comment}

\usepackage{pifont}
\usepackage{tikz}
\usetikzlibrary{positioning,arrows.meta,decorations.pathreplacing,fit}
\tikzset{>=Latex}

\allowdisplaybreaks

\begin{document}

\title{Carbon-aware Market Participation for\\Building Energy Management Systems\\
\thanks{\textsuperscript{*}Authors contributed equally. (Corresponding author: Young-ho Cho.)}
\thanks{This work has been supported by NSF Grants 2130706 and 2150571.}
}

\author{
\IEEEauthorblockN{
Young-ho Cho\textsuperscript{*}, 
Mohamad Chehade\textsuperscript{*}, 
Fatima Al-Janahi\textsuperscript{*}, 
Sol Lim\textsuperscript{*},\\
Javad Mohammadi, and
Hao Zhu
}
\IEEEauthorblockA{
Chandra Family Department of Electrical and Computer Engineering\\
The University of Texas at Austin, Austin, TX, USA \\
\{jacobcho, chehade, FAJ, sollim, javadm, haozhu\}@utexas.edu
}
}

\maketitle

%%%%%%%%%%%%%%%%%%%%%%%%%%%%%%%%%%%%%%%%%%%%%%%%%%%%%%%%%%%%%%%%%%%%%%%%%%%%%%%%%%%%%%%%%%%%%%%%%%%%%%%%%%%%%%%%%

%%%%%%%%%%%%%%%%%%%%%%%%%%%%%%%%%%%%%%%%%%% ABSTRACT %%%%%%%%%%%%%%%%%%%%%%%%%%%%%%%%%%%%%%%%%%% 

\begin{abstract}
Tackling climate change requires the rapid and deep decarbonization of electric power systems. While energy management systems (EMSs) play a central role in this transition, conventional EMSs focus mainly on economic efficiency and often overlook the environmental impact of operational decisions. To address this gap, this paper proposes a unified, real-time building-level carbon-aware EMS (CAEMS) capable of simultaneously co-optimizing grid imports, energy storage, and flexible demand within a single integrated framework. We formulate a mixed-integer linear program (MILP) model that directly integrates time-varying marginal carbon intensity signals into the EMS objective for coordinated participation in both day-ahead (DA) and real-time (RT) markets. To relax the unrealistic assumption of perfect foresight, we incorporate a model predictive control (MPC) extension driven by a Transformer-based forecaster that jointly predicts electricity prices and carbon intensity. The proposed CAEMS is validated using real-world data from the PJM electricity market. Simulation results demonstrate that modest carbon prices can achieve a significant 22.5\% reduction in emissions with only a 1.7\% increase in cost.
\end{abstract}

%\begin{IEEEkeywords}
%Carbon-aware, energy management systems (EMS), model predictive control (MPC), electricity markets
%\end{IEEEkeywords}

%%%%%%%%%%%%%%%%%%%%%%%%%%%%%%%%%%%%%%%%%%%%%%%%%%%%%%%%%%%%%%%%%%%%%%%%%%%%%%%%%%%%%%%%%%%%%%%%%%%%%%%%%%%%%%%%%

%%%%%%%%%%%%%%%%%%%%%%%%%%%%%%%%%%%%%%%%%%% SECTIONS %%%%%%%%%%%%%%%%%%%%%%%%%%%%%%%%%%%%%%%%%%% 

\section{Introduction}
\label{sec:introduction}
\input{sections/introduction_rev.tex}

% \newpage
\section{Carbon-aware Energy Management System}
\label{sec:problem_formulation}
\input{sections/problem_formulation_rev}

% \newpage
\section{Model-Predictive Control (MPC)}
\label{sec:MPC}
\input{sections/MPC_rev}

%\section{Forecasting}
%\label{sec:forecasting}
% \input{sections/Forecasting}

% \newpage
\section{Simulations}
\label{sec:simulations}
\input{sections/simulations_rev}

\section{Conclusion}
\label{sec:conclusion}
\input{sections/conclusion_rev.tex}

%%%%%%%%%%%%%%%%%%%%%%%%%%%%%%%%%%%%%%%%%%%%%%%%%%%%%%%%%%%%%%%%%%%%%%%%%%%%%%%%%%%%%%%%%%%%%%%%%%%%%%%%%%%%%%%%%

%%%%%%%%%%%%%%%%%%%%%%%%%%%%%%%%%%%%%%%%%%% REFERENCES %%%%%%%%%%%%%%%%%%%%%%%%%%%%%%%%%%%%%%%%%%% 
%\newpage
\bibliographystyle{IEEEtran}
\bibliography{ref}

\end{document}

%% file: sections/introduction_rev.tex
Decarbonizing electric power systems has become pivotal in combating climate change. Among various sources of emissions, electricity generation is particularly significant, as the U.S. power sector emitted 1,539 million metric tons of carbon dioxide in 2022, accounting for over 30\% of total energy-related carbon emissions \cite{EIA2023}. Reducing power-sector emissions is therefore critical to achieving global climate objectives and enabling a sustainable transition toward net-zero.

Energy management systems (EMSs) are central to this transition, as they directly control and optimize energy generation, storage operations, and end-use consumption \cite{EMS, EMS2, EMS3, Boost, RLvsMPC}. Traditionally, EMSs have focused on economic efficiency, reliability, and operational optimization. These systems typically aim to minimize costs and maintain system stability, but often overlook the environmental consequences of their operational decisions. As a result, any emission reductions achieved through EMS operation tend to be indirect, arising from efficiency improvements rather than deliberate carbon-aware control \cite{EMSCarbonReduction}. This disconnect limits the alignment of energy system operations with climate goals. As decarbonization becomes a more urgent priority, there is a growing need for EMSs that can co-optimize cost, efficiency, operational control, and carbon emissions.

Carbon-aware concepts have been introduced in the power systems literature in various ways. One stream of work augments EMS formulations with emission-related objectives or constraints, typically through multi-objective optimization that trades off operating cost and carbon emissions \cite{CarbonEMS1, CarbonEMS2, CarbonEMS3}. At the market level, carbon-aware electricity market schemes have been proposed \cite{song2024carbon, chen2024towards}, where locational marginal prices are adjusted using real-time emission data to incentivize consumers to shift demand toward cleaner periods, resulting in measurable reductions in system-wide emissions and improved market efficiency. Concurrently, carbon-aware optimal power flow extends conventional OPF by integrating carbon emission flow modeling into dispatch decisions, thereby co-optimizing power flows and associated emissions across the grid \cite{shao2010co, chen2024carbon}. Furthermore, carbon-aware demand response (DR) programs have been developed across various sectors, leveraging real-time price and carbon signals to achieve simultaneous cost savings and emissions reductions \cite{chen2024enhance, lee2025joint, wang2021optimal}. Recent work has applied such carbon-aware DR to residential and commercial buildings, demonstrating substantial reductions in both operating costs and carbon emissions \cite{Yang2025CarbonAwareSO}. While these efforts have advanced carbon-aware operation within specific domains of the power system, most approaches address individual components independently. Consequently, a unified, real-time carbon-aware energy management system (CAEMS) capable of simultaneously co-optimizing generation and flexible load within a single integrated framework remains unexplored.

To address this, we put forth a unified CAEMS, designed to integrate carbon intensity signals directly into the building-level EMS optimization problem. Unlike conventional approaches that emphasize supply-side carbon mitigation strategies, the proposed framework treats demand-side flexibility and energy storage as active, first-class resources. It enables coordinated control across day-ahead (DA) and real-time (RT) electricity markets, co-optimizing for both cost and emissions.  As a result, energy consumption and storage schedules can be dynamically adjusted to align with periods of lower marginal emissions and prices, without compromising reliability. This approach delivers immediate emissions reductions by turning market signals into actionable control decisions, integrating environmental awareness into routine market participation, and supporting steady progress toward long-term sustainability goals. The main contributions can be summarized as follows:
\begin{enumerate}
    \item We propose a unified optimization framework that co-optimizes grid imports, energy storage, and flexible demand under both cost and carbon reduction objectives.
    \item We formulate a mixed-integer linear program (MILP)-based optimization model that enables coordinated control across both DA and RT electricity markets.
    \item We introduce a Transformer-based forecasting module for joint prediction of electricity prices and carbon intensity, enabling an MPC extension of the proposed CAEMS that can operate without perfect foresight.
\end{enumerate}

%% file: sections/problem_formulation_rev.tex
The carbon-aware energy management system (CAEMS) is a building-level control framework that schedules real-time flexibility to reduce operating cost and carbon emissions operating around a given day-ahead (DA) schedule. The CAEMS is formulated over a multi-period horizon and determines real-time (RT) dispatch decisions for the grid import, the energy storage system (ESS), and flexible demand. By explicitly incorporating carbon intensity signals in addition to electricity prices, the CAEMS enables the building to respond to both economic and environmental incentives while respecting a pre-committed DA baseline.

We formulate the CAEMS as a finite-horizon MILP whose decision variables capture real-time grid imports, ESS operation, and flexible demand, while treating DA schedules as fixed parameters. The DA schedule $\beta_t$ is determined by a higher-level scheduling process and represents the financial baseline position in the DA market. At each time $t$, the decision variable $g_t$ denotes the net real-time power imported from the grid, while the ESS operation and flexible demand are described by power and state variables such as $g^{b}_t$, $b^{g}_t$, $b_t$, $\mathrm{SoC}_t$, and $\Delta_t$. Exogenous signals include the inelastic demand $d_t$, DA and RT prices $\pi^{\mathrm{DA}}_t$, $\pi^{\mathrm{RT}}_t$, and the corresponding carbon intensities $c^{\mathrm{DA}}_t$, $c^{\mathrm{RT}}_t$. In addition, binary variables $u^{\mathrm{ch}}_t$ and $u^{\mathrm{dis}}_t$ indicate charging and discharging modes of the ESS and prevent simultaneous charging and discharging.

The CAEMS objective is to minimize the sum of monetary energy cost and carbon cost incurred by DA and RT market settlements. We write the total cost as
\begin{equation}
    J = p^E + p^C,
\end{equation}
where $p^E$ is the energy cost and $p^C$ is the carbon cost. The energy cost aggregates DA and RT settlements as
\begin{equation}
    p^E = \sum_{t=1}^{T} \left( \pi^{\mathrm{DA}}_t \beta_t + \pi^{\mathrm{RT}}_t (g_t - \beta_t) \right),
\end{equation}
with $\pi^{\mathrm{DA}}_t$ and $\pi^{\mathrm{RT}}_t$ denoting the DA and RT electricity prices. Similarly, the carbon cost accounts for emissions attributed to DA and RT energy:
\begin{equation}
    p^C = \sum_{t=1}^{T} \left( \pi^{C} c^{\mathrm{DA}}_t \beta_t + \pi^{C} c^{\mathrm{RT}}_t (g_t - \beta_t) \right),
\end{equation}
where $c^{\mathrm{DA}}_t$ and $c^{\mathrm{RT}}_t$ are the DA and RT marginal carbon intensities, and $\pi^{C}$ is the carbon price (or tax rate) per emission unit. Since $\beta_t$ is fixed, the DA-related terms in the above expressions are constant with respect to the decision variables and can be omitted in numerical implementation, but are retained here to clearly distinguish DA and RT contributions. The resulting optimization problem is written as
\begin{equation}
    \min_{\{g_t,g^{b}_t,b^{g}_t,b_t,\Delta_t,u^{\mathrm{ch}}_t,u^{\mathrm{dis}}_t,\ldots\}} \; p^E + p^C.
\end{equation}

This optimization is subject to physical and operational constraints that capture the building power balance, ESS dynamics, and limits on flexible demand. The building-level energy balance ensures that inelastic demand and shifted demand are supplied by the grid and ESS. This relationship is expressed as
\begin{equation}
    d_t = g^{d}_t + b^{d}_t + (\Delta_t - \Delta_{t-1}),
\end{equation}
where $g^{d}_t$ and $b^{d}_t$ denote the portions of demand served by the grid and ESS, and $\Delta_t$ represents cumulative demand shifting. The net grid import is defined as
\begin{equation}
    g_t = g^{d}_t + g^{b}_t - b^{g}_t,
\end{equation}
with $g^{b}_t$ and $b^{g}_t$ capturing ESS charging and discharging power exchanged with the grid. ESS power and state-of-charge (SoC) dynamics are modeled by
\begin{equation}
    b_t = g^{b}_t - b^{g}_t,
\end{equation}
\begin{equation}
    \mathrm{SoC}_{t+1} = \mathrm{SoC}_{t} + b_t,
\end{equation}
together with SoC and power limits
\begin{equation}
    \mathrm{SoC}^{\min} \leq \mathrm{SoC}_t \leq \mathrm{SoC}^{\max}, \quad \forall t,
\end{equation}
\begin{equation}
    0 \le g^{b}_t \le a^{\max}, \quad 0 \le b^{g}_t \le a^{\max}, \quad \forall t,
\end{equation}
where $a^{\max}$ denotes the maximum charge/discharge rate. The binary variables enforce mutually exclusive charging and discharging modes:
\begin{equation}
    u^{\mathrm{ch}}_t, u^{\mathrm{dis}}_t \in \{0,1\}, \quad \forall t,
\end{equation}
\begin{equation}
    0 \le g^{b}_t \le u^{\mathrm{ch}}_t a^{\max}, \quad 0 \le b^{g}_t \le u^{\mathrm{dis}}_t a^{\max}, \quad \forall t,
\end{equation}
\begin{equation}
    u^{\mathrm{ch}}_t + u^{\mathrm{dis}}_t \le 1, \quad \forall t.
\end{equation}
Flexible demand is bounded and enforced to return to its baseline by the end of the horizon:
\begin{equation}
    0 \leq \Delta_t \leq \Delta^{\max}, \quad \forall t,
\end{equation}
\begin{equation}
    \Delta_0 = \delta, \quad \Delta_T = 0,
\end{equation}
where $\Delta^{\max}$ is the maximum allowable cumulative shift and $\delta$ is the initial offset.

Together, the objective and constraints define a deterministic multi-period MILP that optimizes real-time operation around a given DA baseline and is later embedded in a model predictive control (MPC) framework to handle forecast uncertainty. The deterministic CAEMS model in the above equations assumes perfect knowledge of future prices and carbon intensities. In practice, these exogenous signals are only available through forecasts; the next section therefore uses MPC to repeatedly solve this optimization with updated predictions and implement only the first-step decision at each control interval.

%% file: sections/MPC_rev.tex
Traditional day-ahead and real-time optimization assumes perfect foresight of prices, loads, and emission intensities, which can lead to suboptimal or infeasible schedules when actual conditions deviate from forecasts. To address this, we adopt a MPC approach that continuously updates decisions in a receding-horizon framework. At each control interval $t$, the EMS solves a finite-horizon mixed-integer program over $\{t,\ldots,t+H-1\}$, implements only the first-step decision, then incorporates newly observed data and updated forecasts before re-solving.

In the MPC formulation, real-time price and emission-intensity parameters are replaced by time-indexed forecasts via estimators $f_p$ and $f_c$, respectively, which are parameterized by the \emph{state of information} $\theta_t$. For notational convenience, we also define carbon-cost coefficients (in \$/MWh) as $\lambda^{\mathrm{DA}}_{t} \triangleq \pi^{C} c^{\mathrm{DA}}_{t}$ and $\hat{\lambda}^{\mathrm{RT}}_{t+k\mid t} \triangleq \pi^{C} f_c(t+k\mid t;\theta_t)$; if day-ahead marginal emission rates are not available, $\lambda^{\mathrm{DA}}_{t}$ can be formed from an exogenous hourly marginal-emissions signal (e.g., an aggregation of real-time rates) or replaced by the corresponding forecast without changing the MPC structure.

\begin{align}
\min \quad & p^{E}_{t} + p^{C}_{t} \label{eq:mpc_obj} \\[2pt]
p^{E}_{t} \;=\; & \sum_{k=0}^{H-1} \Big(
\pi^{\mathrm{DA}}_{t+k}\, \beta_{t+k}
\;+\; f_p(t+k\mid t;\theta_t)\,\big(g_{t+k\mid t}-\beta_{t+k}\big)
\Big), \label{eq:mpc_energy} \\[2pt]
p^{C}_{t} \;=\; & \sum_{k=0}^{H-1} \Big(
\lambda^{\mathrm{DA}}_{t+k}\, \beta_{t+k}
\;+\; \hat{\lambda}^{\mathrm{RT}}_{t+k\mid t}\,\big(g_{t+k\mid t}-\beta_{t+k}\big)
\Big). \label{eq:mpc_carbon}
\end{align}

The MPC problem is subject to the same physical and market participation constraints as in Section~\ref{sec:problem_formulation}, with the current state (e.g., $\mathrm{SOC}_{t}$ and any carried-over flexibility state) fixed to its most recent value. Only the first-step decisions (e.g., $\{g_{t\mid t}, b_{t\mid t}, \Delta_{t\mid t}\}$) are implemented; the remaining planned actions are discarded once new measurements become available.

\subsection{Transformer-Based Forecasting Estimator}

Accurate multi-step forecasts of day-ahead prices $\pi^{\mathrm{DA}}_t$, real-time prices $\pi^{\mathrm{RT}}_t$, demand $d_t$, and emission intensities $c^{\mathrm{DA}}_t, c^{\mathrm{RT}}_t$ are critical for MPC performance. We therefore employ a Transformer network, a deep learning architecture built exclusively on self-attention, which excels at capturing long-range temporal dependencies without recurrence~\cite{attention}. The model takes as input a historical sequence
\begin{equation}
    \{ x_{t-\tau}, \ldots, x_{t-1} \}, \quad
    x_t = [ \pi^{\mathrm{DA}}_t, \pi^{\mathrm{RT}}_t, d_t, c^{\mathrm{DA}}_t, c^{\mathrm{RT}}_t ],
\end{equation}
to predict the $H$-step-ahead sequence $\{ \hat{x}_{t\mid t}, \hat{x}_{t+1\mid t}, \ldots, \hat{x}_{t+H-1\mid t} \}$.
Sinusoidal positional encodings are first added to each input embedding to encode time-step order. Each Transformer layer then applies a multi-head self-attention mechanism---where scaled dot-product attention computes weights between all pairs of positions---followed by a position-wise feed-forward network consisting of two linear transformations with a non-linear activation. Residual connections and layer normalization follow each sublayer to stabilize training and improve convergence. We train the network end-to-end by minimizing the mean squared error between predicted and actual values over a rolling-window dataset. At runtime, the trained Transformer supplies the MPC with the forecasts $\hat{\pi}^{\mathrm{RT}}_{t+k\mid t}$ and $\hat{c}^{\mathrm{RT}}_{t+k\mid t}$ over $k=0,\ldots,H-1$ at each control interval, enabling robust, carbon-aware dispatch under uncertainty.

%% file: sections/simulations_rev.tex
% \begin{figure}[t!]
%   \centering
%   \includegraphics[width=0.6\linewidth]{figures/price}
%   \caption[]{\small Hourly electricity prices (\$/MWh) of DA and RT markets.}
%   \label{fig:price_ts}
% \end{figure}

% \begin{figure}[t!]
%   \centering
%   \includegraphics[width=0.6\linewidth]{figures/rate}
%   \caption[]{\small Hourly marginal emission rates (kg CO$_2$/MWh) of DA and RT markets.}
%   \label{fig:input_data}
% \end{figure}

\begin{figure}[t!]
    \centering
    \begin{tabular}[b]{@{}c@{}}
        \includegraphics[width=0.47\linewidth]{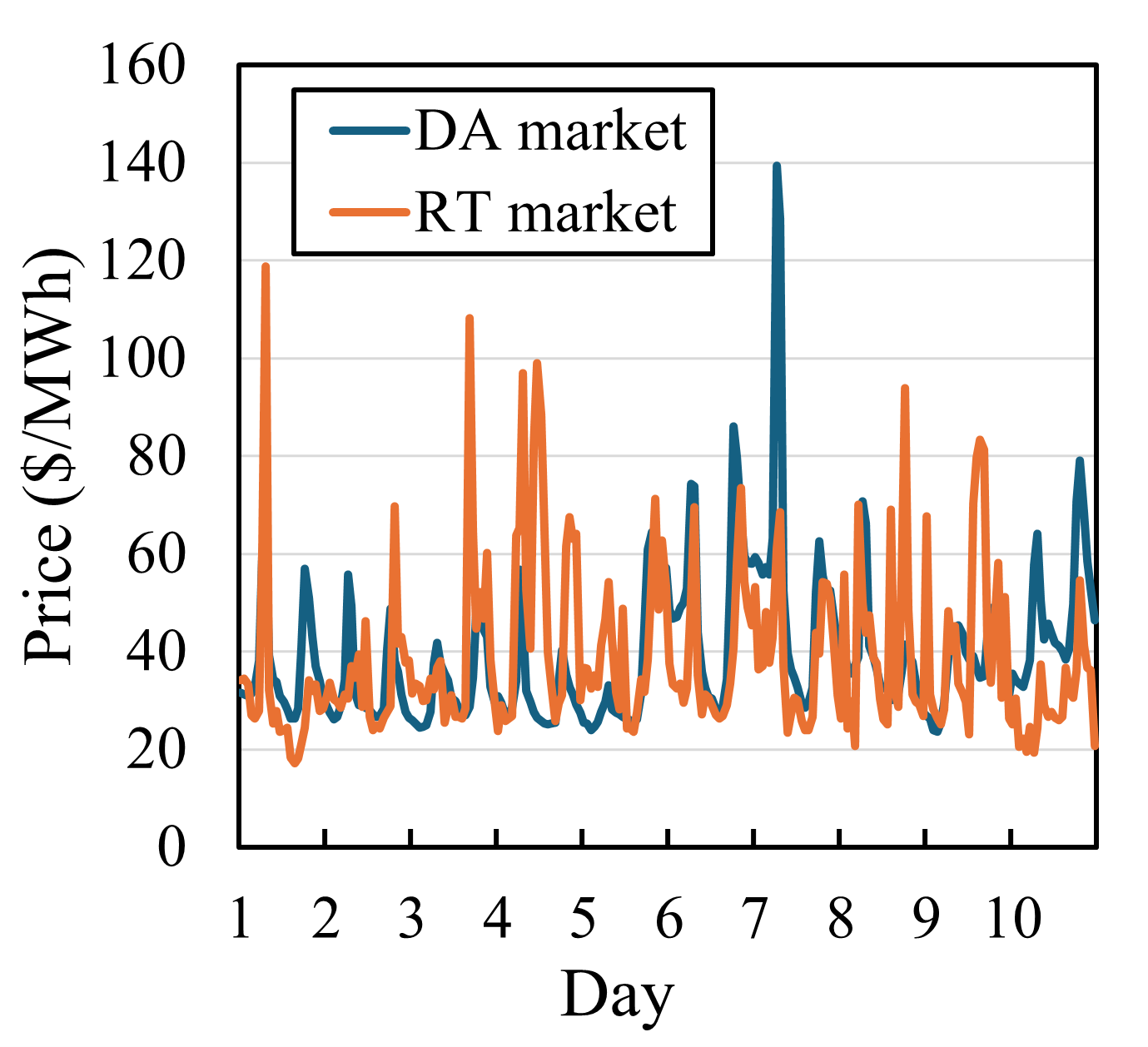} \\
        \small (a) \\
    \end{tabular}\quad
    \begin{tabular}[b]{@{}c@{}}
        \includegraphics[width=0.47\linewidth]{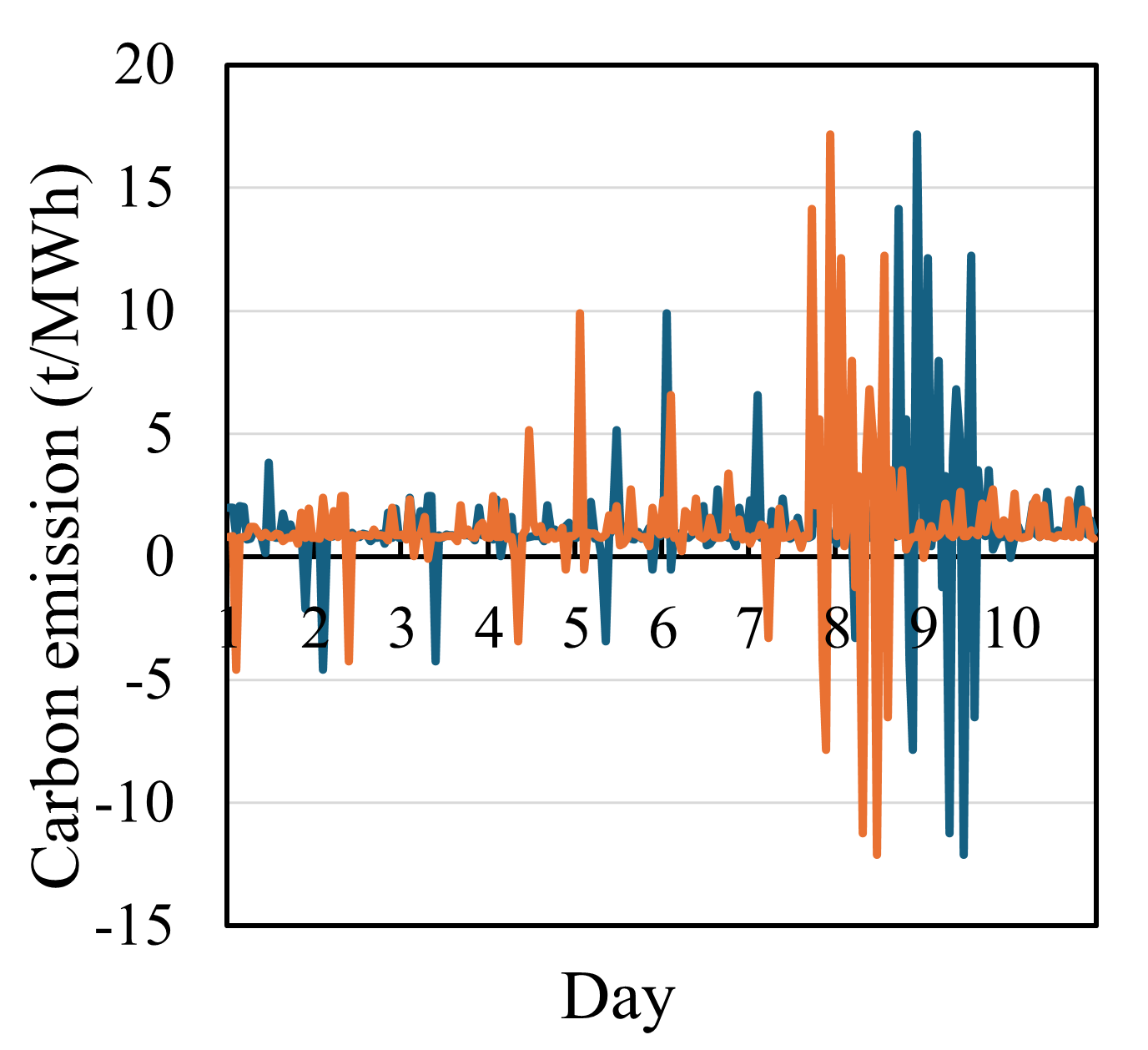} \\
        \small (b) \\
    \end{tabular}
    \caption{PJM datasets of DA and RT markets: (a) Hourly electricity prices (\$/MWh) (b) Hourly marginal emission rates (t CO$_2$/MWh).}
    \label{fig:input_data}
\end{figure}

\begin{figure*}[!t]
\centering
\resizebox{\textwidth}{!}{%
\begin{tikzpicture}[
  every node/.style={font=\footnotesize},
  box/.style={draw, rounded corners=1mm, fill=black!5, inner sep=4pt, align=left},
  core/.style={draw, rounded corners=1mm, fill=black!10, inner sep=6pt, minimum width=36mm, minimum height=13mm, align=center},
  lite/.style={draw, rounded corners=0.8mm, inner sep=2pt, align=left},
  arr/.style={->, line width=0.8pt},
  node distance=24mm
]

% ---------- Left block: data prep ----------
\node[box] (past) {\textbf{Inputs (last 24 h)}\\
\(\mathbf{p}_{t-23:t}\in\mathbb{R}^{24\times 1}\): RT price [\$/MWh]\\
\(\mathbf{c}_{t-23:t}\in\mathbb{R}^{24\times 1}\): carbon intensity [kg CO\(_2\)/MWh]};

% small note about alignment + ffill (from text)
\node[lite, below=2mm of past, align=left] (prep) {aligned to hourly grid; missing values \emph{forward-filled}};

% ---------- Forecaster ----------
\node[core, right=28mm of past] (f) {\textbf{Forecaster} \(f\) \\(Transformer selected by RMSE, Tab.~\ref{tab:forecasting_results})};

% ---------- Right block: outputs ----------
\node[box,  right=28mm of f] (future) {\textbf{Outputs (next 24 h)}\\
\(\hat{\mathbf{p}}_{t+1:t+24}\in\mathbb{R}^{24\times 1}\): RT price forecast\\
\(\hat{\mathbf{c}}_{t+1:t+24}\in\mathbb{R}^{24\times 1}\): carbon intensity forecast};

% ---------- Arrows with labels ----------
\draw[arr] (past.east) -- node[above,pos=.52]{\(\mathbf{p}_{t-23:t}\)} (f.west);
\draw[arr, yshift=-4pt] (past.east) -- node[below,pos=.52]{\(\mathbf{c}_{t-23:t}\)} (f.west);

\draw[arr] (f.east) -- node[above,pos=.48]{\(\hat{\mathbf{p}}_{t+1:t+24}\)} (future.west);
\draw[arr, yshift=-4pt] (f.east) -- node[below,pos=.48]{\(\hat{\mathbf{c}}_{t+1:t+24}\)} (future.west);

% ---------- Braces for window & horizon ----------
\draw[decorate,decoration={brace,amplitude=3.5pt},yshift=6pt]
  (past.north west) -- node[above=5pt]{\textit{window } \(t-23{:}t\)} (past.north east);
\draw[decorate,decoration={brace,amplitude=3.5pt},yshift=6pt]
  (future.north west) -- node[above=5pt]{\textit{horizon } \(t{+}1{:}t{+}24\)} (future.north east);

% ---------- Functional mapping (compact, centered) ----------
\node[below=12mm of f, align=center] (map)
  {\(f:\;(\mathbf{p}_{t-23:t},\mathbf{c}_{t-23:t})\mapsto(\hat{\mathbf{p}}_{t+1:t+24},\hat{\mathbf{c}}_{t+1:t+24})\)};

% Optional thin grouping boxes (no fill; comment out if you want ultra-minimal)
\node[fit=(past)(prep), draw, rounded corners=1mm, inner sep=3pt] {};
\end{tikzpicture}%
}
\caption[]{\small Two-series, 24-hour-ahead forecasting setup. The model \(f\) consumes the last 24 hours of \emph{real-time} (RT) price \(\mathbf{p}\) and carbon intensity \(\mathbf{c}\) (aligned hourly; forward-filled if missing) and outputs 24-hour forecasts \(\hat{\mathbf{p}},\hat{\mathbf{c}}\). The forecaster is implemented with a Transformer chosen by RMSE (Table~\ref{tab:forecasting_results}).}
\label{fig:forecast-arch-wide}
\end{figure*}

The proposed carbon-aware EMS is evaluated using real-world hourly locational marginal prices (LMPs) and marginal carbon emission rates obtained from PJM \cite{pjm_dataminer2}. We focus on a 10-day period to capture a range of market conditions. The input dataset comprises hourly day-ahead (DA) and real-time (RT) prices (see Fig.~\ref{fig:input_data}(a)) alongside corresponding marginal emission rates in t CO$_2$/MWh (see Fig.~\ref{fig:input_data}(b)). All time series were aligned to a common hourly grid, and any missing observations were forward-filled to ensure continuity.

To validate the proposed EMS, we establish a baseline representing an unmanaged operation where grid imports exactly match the building demand without leveraging storage or flexibility resources. Against this baseline, we implement two EMS-based dispatch strategies: the \emph{cost-only dispatch} minimizes energy procurement cost only (equivalent to a carbon price of \$0/t), whereas the \emph{carbon-aware dispatch} adds a linear carbon penalty of \$30/t to the objective. For each day in the dataset, the EMS solves the MILP described in Section II over a 24~h rolling horizon with perfect foresight. The optimization determines hourly grid imports, battery charge/discharge schedules, and demand-shift actions, all subject to the operational and technical constraints previously defined.

Performance is assessed in terms of total daily energy cost, aggregate carbon emissions, and the intraday profiles of grid imports and emissions. This simulation framework enables a direct evaluation of the cost and emissions impacts of incorporating a carbon penalty into participation decisions.

\subsection{Forecasting model Validation}
\label{sec:forecaster}
First, we design a forecaster $f$ to jointly predict the \textbf{electricity prices} and \textbf{carbon emission intensity}, as shown in Fig.~\ref{fig:forecast-arch-wide}. We consider real data from PJM \cite{pjm_dataminer2}, and test several predictive models constructed with supervised learning. The forecasting results are shown in Table \ref{tab:forecasting_results}, where the transformer model is shown to be the best. Hence, we use one transformer model to jointly forecast the real-time market clearing price and carbon emission intensity. 

Although Table~\ref{tab:forecasting_results} reports results on PJM data, the forecasting module is used as a plug-in that consumes generic market time series (prices and marginal emissions), so differences in market design, regulation, or renewable penetration primarily enter through the statistics of these exogenous signals rather than requiring changes to the CAEMS formulation. Furthermore, the receding-horizon nature of the MPC inherently mitigates forecast errors. By re-optimizing with updated measurements at every control interval, the system dynamically corrects for prediction mismatches, ensuring robust dispatch performance.
%Moreover, because the EMS is operated in a receding-horizon MPC fashion, solving with newly observed data and applying only the first-step decision. Therefore, forecast errors are corrected online, which helps limit the impact of prediction mismatch on realized dispatch.

% \begin{figure}[t!]
%     \centering
%     \includegraphics[width=0.8\linewidth]{figures/forecaster}
%     \caption{The framework for the forecaster. The forecaster can be, but is not limited to, supervised learning predictive models.}
%     \label{fig:forecaster}
% \end{figure}

% Preamble (once):
% \usepackage{tikz}
% \usetikzlibrary{positioning,arrows.meta,decorations.pathreplacing,fit}
% \tikzset{>=Latex}

\begin{table}[t]
\centering
\caption{Root Mean Squared Error (RMSE) of forecasting models on PJM data. Models are sorted by performance, and the Transformer model achieves the lowest RMSE, indicating the best predictive accuracy.}
\label{tab:forecasting_results}
\begin{tabular}{ll|ll}
\toprule
\textbf{Model} & \textbf{RMSE} & \textbf{Model} & \textbf{RMSE} \\
\midrule
\textbf{Transformer}      & \textbf{0.0462} & XGBoost          & 0.0528 \\
CatBoost         & 0.0469 & LSTM             & 0.0566 \\
LightGBM         & 0.0471 & Linear Regression& 0.0569 \\
Random Forest    & 0.0473 & TCN              & 0.0573 \\
AdaBoost         & 0.0476 & Decision Tree    & 0.0631 \\
KNN              & 0.0480 & SVM              & 0.0722 \\
MLP              & 0.0484 & ARIMA            & 0.0816 \\
\bottomrule
\end{tabular}
\end{table}

\subsection{Optimization results}
Simulations over the 10‑day dataset assess the impact of a linear carbon tax on economic dispatch. We evaluate four tax levels—\$0, \$10, \$20, and \$30/t—plus a no‑optimization baseline. Each scenario is solved as an MILP with perfect foresight in both DA and RT markets.

\begin{figure}[t!]
  \centering
  \includegraphics[width=0.35\textwidth]{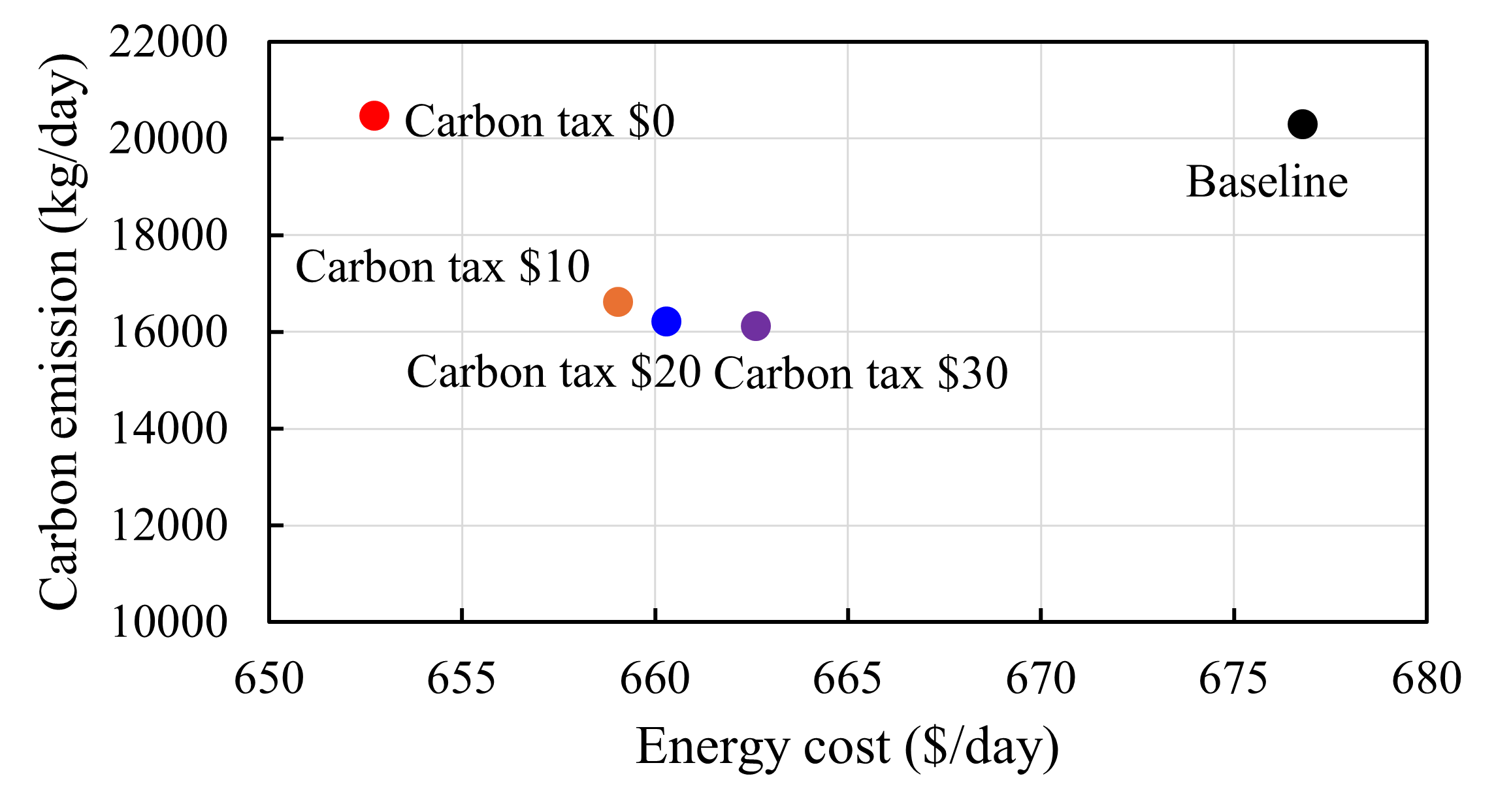}
  \caption[]{\small Daily trade‑off between energy cost and carbon emissions under different carbon tax scenarios.}
  \label{fig:cost_emissions}
\end{figure}

Fig.~\ref{fig:cost_emissions} shows the steady‑state trade‑off between daily energy cost and carbon emissions. At zero tax, the dispatch achieves the lowest cost (\$653/day) but the highest emissions (20.4 t/day). As the tax increases to \$10, \$20, and \$30/t, emissions decrease to approximately 16.5–15.8 t/day, while cost rises modestly to about \$664/day. A comparison with the baseline highlights that cost-only dispatch even increases emissions by ignoring environmental impact. In contrast, the carbon-aware dispatch effectively reduces both energy costs and carbon emissions in an efficient manner.

%\textcolor{red}{The “No optimization” marker demonstrates that even a minimal carbon price outperforms a cost‑only benchmark in both economic and environmental metrics.}

\begin{figure*}[t!]
    \centering
    \begin{tabular}[b]{@{}c@{}}
        \includegraphics[width=0.32\textwidth]{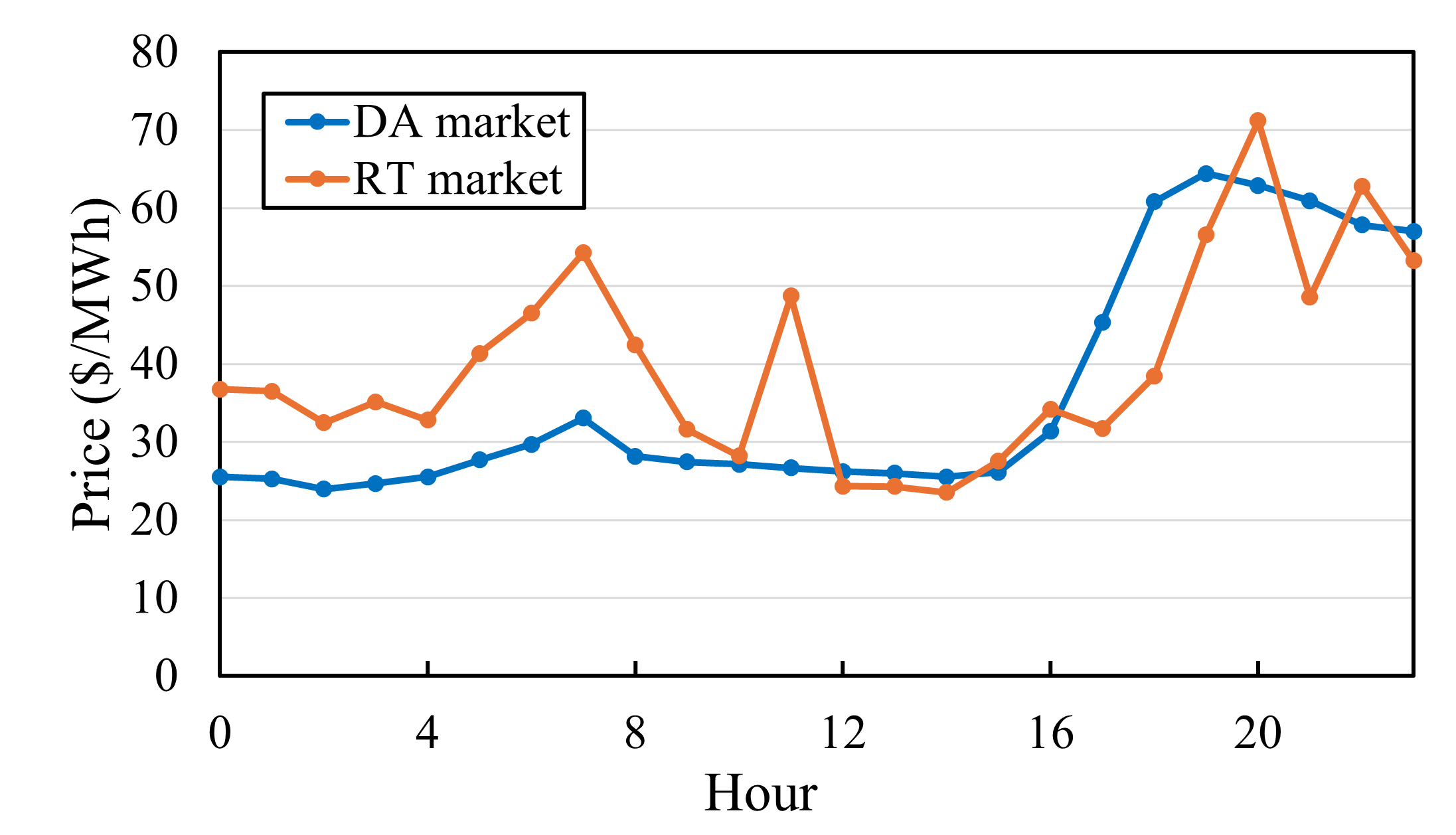} \\
        \small (a) \\
    \end{tabular}
    \hfill
    \begin{tabular}[b]{@{}c@{}}
        \includegraphics[width=0.32\textwidth]{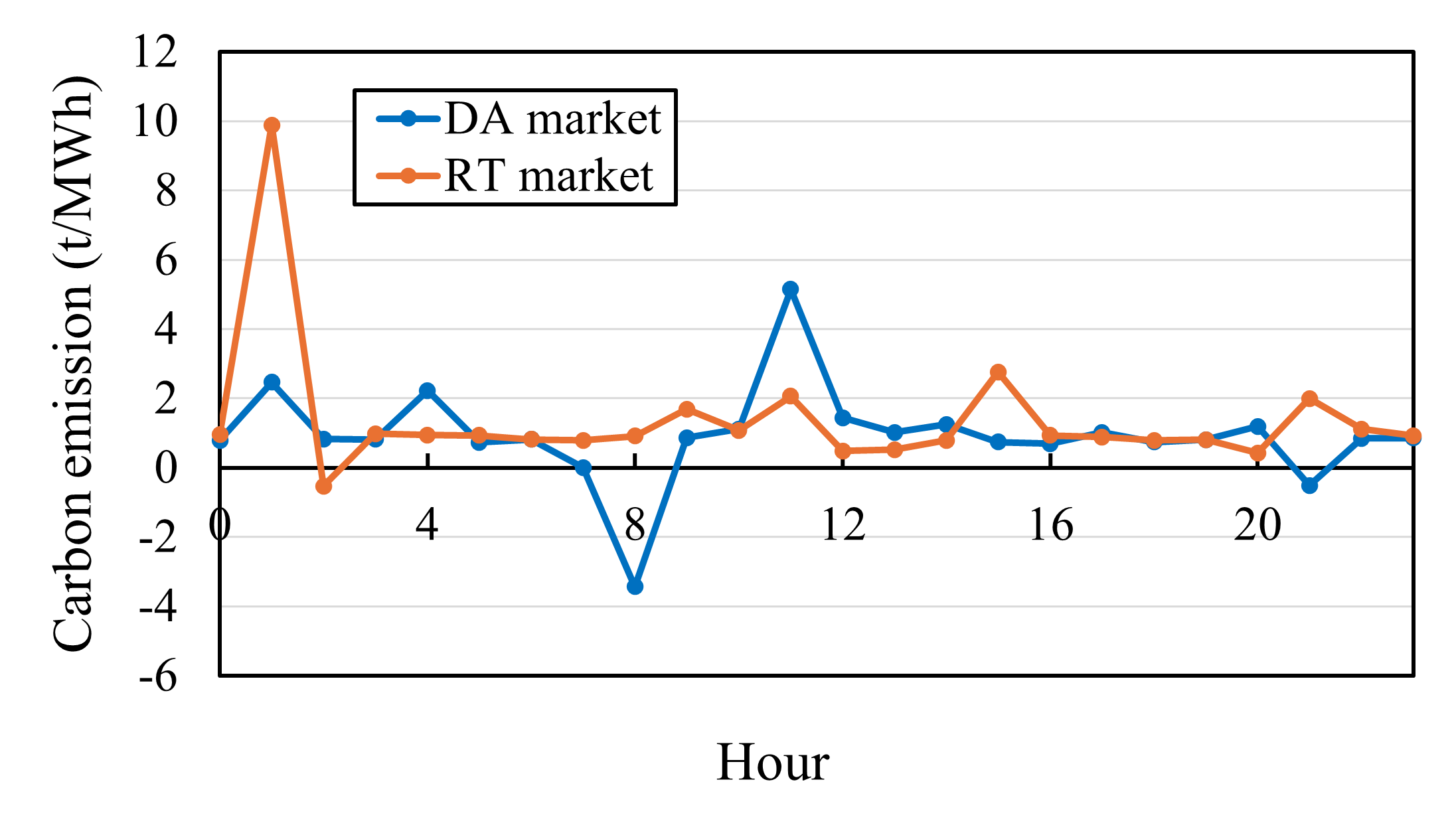} \\
        \small (b) \\
    \end{tabular}
    \hfill
    \begin{tabular}[b]{@{}c@{}}
        \includegraphics[width=0.32\textwidth]{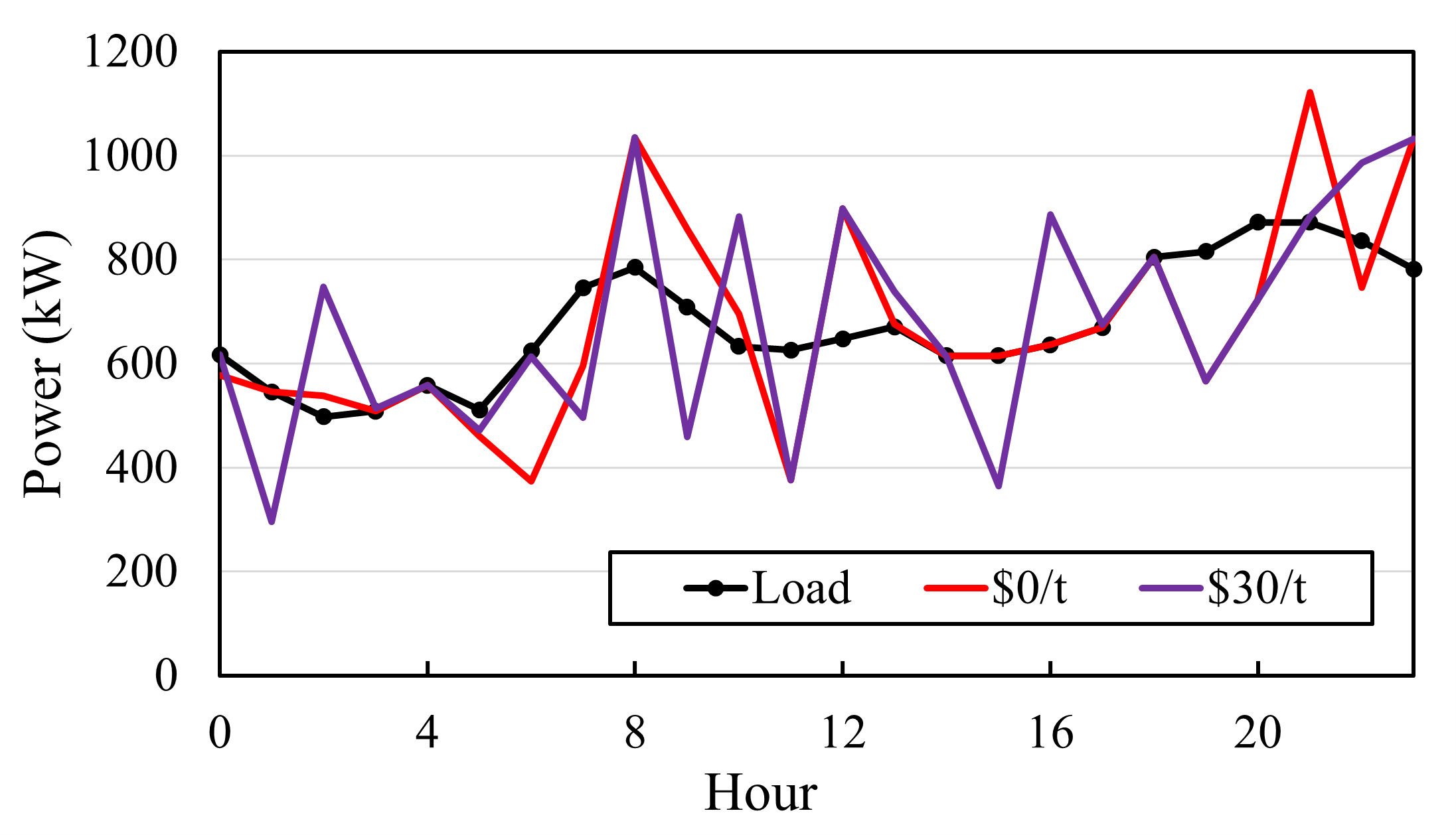} \\
        \small (c) \\
    \end{tabular}
    \caption{Intraday dynamics: (a) Hourly electricity prices in DA and RT markets for a single day. (b) Hourly emission rates in DA and RT markets for a single day. (c) Comparison of optimized net grid imports for the cost-only (\$0/t) and carbon-aware (\$30/t) strategies.}
    \label{fig:intraday dynamics}
\end{figure*} 

To illustrate intraday dynamics, we compare the \$0 and \$30/t cases for a representative day. Fig.~\ref{fig:intraday dynamics}(a) and \ref{fig:intraday dynamics}(b) plot hourly electricity prices and marginal emission rates in DA and RT markets, respectively. Fig.~\ref{fig:intraday dynamics}(c) compares the net grid imports, revealing that the carbon-aware dispatch significantly alters the procurement schedule to avoid high-emission periods. Both strategies reduce grid imports during the midday price spike. However, only the carbon‑aware dispatch mitigates the early‑morning emission peak by curtailing imports. This confirms that the \$30/t model strategically cuts purchases during high‑carbon hours.
Overall, these results confirm that embedding a carbon price within the EMS delivers dual benefits: (i) cost-effective mitigation of price spikes and (ii) targeted emission reductions during critical hours—capabilities unattainable by a traditional cost-only dispatch.

% \begin{figure}[t]
%   \centering
%   \includegraphics[width=0.35\textwidth]{figures/oneday_price.png}
%   \caption[]{\small Hourly electricity prices in DA and RT markets for a single day.}
%   \label{fig:hourly_analysis_price}
% \end{figure}

% \begin{figure}[t]
%   \centering
%   \includegraphics[width=0.35\textwidth]{figures/oneday_rate.png}
%   \caption[]{\small Hourly emission rates in DA and RT markets for a single day.}
%   \label{fig:hourly_analysis_rate}
% \end{figure}

% \begin{figure}[t]
%   \centering
%   \includegraphics[width=0.35\textwidth]{figures/oneday.png}
%   \caption[]{\small Comparison of optimized net grid imports for the cost-only (\$0/t) and carbon-aware (\$30/t) strategies.}
%   \label{fig:hourly_analysis_opt}
% \end{figure}

However, introducing a non-zero carbon price may generate customer resistance and raise regulatory considerations, potentially affecting market acceptance and policy implementation.
From a customer perspective, this trade-off is highly favorable: a marginal 1.7\% increase (from \$653 to \$664) in daily energy expenditure yields a substantial 22.5\% reduction (from 20.4 t to 15.8 t) in carbon emissions.
Although such rate adjustments can prompt rate-case proceedings and customer inquiries, experience shows that incremental increases below 2\% typically fall within accepted tolerance thresholds for both residential and commercial consumers~\cite{eia2024electricitybills}. Moreover, participants in demand-response (“negawatt”) programs can receive incentive payments that offset much of this additional cost, effectively transforming a potential burden into a revenue stream and enhancing program engagement.

For the market operator, integrating carbon pricing into dispatch and settlement processes necessitates enhancements to market-clearing algorithms, real-time settlement infrastructure, and compliance reporting frameworks to incorporate locational marginal carbon emission (LMCE) signals. While these upgrades introduce additional computational overhead and require close coordination among stakeholders, the resulting operational shifts—such as flattened peak imports, more evenly distributed net-import profiles, and reduced dependence on fossil-heavy generation—yield a measurably cleaner grid, improved system reliability, and stronger alignment with state and federal decarbonization mandates. By coupling advanced AI-driven forecasting with optimization-based EMS, PJM can implement these mechanisms with minimal disruption, ensuring that carbon costs are transparently valued without compromising market liquidity or affordability.

%% file: sections/conclusion_rev.tex
This work demonstrates that carbon-aware dispatch is both technically feasible and economically viable, charting a clear pathway for PJM to achieve its sustainability and reliability objectives while providing actionable insights for policymakers, market operators, and end users.

While our simulations use a linear per-ton carbon price to illustrate the core efficacy of the CAEMS, the proposed MILP is readily extendable to more complex policy mechanisms. For example, cap-and-trade mechanisms can be incorporated by introducing emissions-budget constraints, while offset-credit programs can be modeled through policy-dependent carbon price signals; tiered pricing can be represented through piecewise-linear cost terms. These extensions would further enable demand-side carbon-aware operation as a practical pathway toward power system decarbonization.